\documentclass[runningheads]{llncs}
\usepackage{graphicx}
\usepackage[hyphens]{url}
\usepackage{hyperref}
\usepackage{color}

\usepackage[font=small,labelfont=bf]{caption} 
\usepackage[subrefformat=parens]{subcaption}

\usepackage{algorithm}
\usepackage{algorithmic}
\usepackage{multirow}
\usepackage{amssymb}
\usepackage{amsmath}

\DeclareMathOperator*{\argmin}{arg\,min}

\begin{document}
\title{Unfolded Self-Reconstruction LSH: Towards Machine Unlearning in Approximate Nearest Neighbour Search}
%
%


\author{
Kim Yong Tan\inst{1} \and
Yueming Lyu\inst{2}\thanks{Corresponding Author} \and
Yew Soon Ong\inst{1} \and
Ivor W. Tsang\inst{2}
}
\authorrunning{KY. Tan, Y. Lyu, et al.}
%

\institute{
Nanyang Technological University, Singapore \\ \email{\{kimyong001,asysong\}@ntu.edu.sg} \and 
Centre for Frontier AI Research (CFAR) \\
Agency for Science, Technology and Research (A*STAR) \\ \email{\{lyu\_yueming,ivor\_tsang\}@cfar.a-star.edu.sg}
}

\maketitle              
\begin{abstract}

Approximate nearest neighbour (ANN) search is an essential component of search engines, recommendation systems, etc.   Many recent works focus on learning-based data-distribution-dependent hashing and achieve good retrieval performance. However,  due to increasing demand for users' privacy and security, we often need to remove users' data information from Machine Learning (ML) models to satisfy specific privacy and security requirements \cite{bourtoule2021machine,voigt2017eu,goldman2020introduction,privacy}. This need requires the ANN search algorithm to support fast online data deletion and insertion. Current learning-based hashing methods need retraining the hash function, which is prohibitable due to the vast time-cost of large-scale data. To address this problem, we propose a novel data-dependent hashing method named \textbf{unfolded self-reconstruction locality-sensitive hashing (USR-LSH)}. Our USR-LSH unfolded the optimization update for instance-wise data reconstruction, which is better for preserving data information than data-independent LSH. Moreover, our USR-LSH supports fast online data deletion and insertion without retraining. 
To the best of our knowledge, we are the first to address the machine unlearning of retrieval problems. Empirically, we demonstrate that USR-LSH outperforms the state-of-the-art data-distribution-independent LSH in ANN tasks in terms of precision and recall. We also show that USR-LSH has significantly faster data deletion and insertion time than learning-based data-dependent hashing.

\end{abstract}
\section{Introduction}

Given a collection of data points $\boldsymbol{X} = \{x_1,x_2, ...,x_N\}$, and a query point $\boldsymbol{q}$, an exact k nearest neighbour (k-NN) search problem is to find k distinct points that are nearest to the query point $\boldsymbol{q}$. There exist efficient algorithms \cite{bentley1975multidimensional} for exact nearest neighbour search in low-dimensional cases. In large-scale high-dimensional cases, it turns out that the problems become hard, and most algorithms even take higher computational costs than the linear scan \cite{wang2014hashing}. This phenomenon is often called the "curse of dimensionality" \cite{indyk1998approximate}. Therefore, a lot of recent efforts are focused on searching for approximate nearest neighbours (ANN) \cite{indyk1998approximate}. The goal is to find $k$ distinct points that are "close enough" to the given query point $\boldsymbol{q}$ with a mild trade-off of accuracy for a significant improvement in efficiency.


One of the popular algorithms to solve the ANN search problem is Locality-sensitive hashing (LSH) \cite{indyk1998approximate,gionis1999similarity}. The idea is to hash similar data points to the same hash code with high probability. Binary LSH is one kind of LSH that generates binary hash codes. It can approximate the Euclidean or angular distance between data points by the hamming distance between their corresponding hash codes. Computing the hamming distance is faster than computing other distances, such as Euclidean and cosine distance, because it mainly involves bitwise operations. Eight bits of binary hash codes can be packed into one byte, and due to its compact bit representation, it can reduce the storage requirement. However, the performance of data-independent LSH is limited because it is totally probabilistic, i.e., the hash function is independent of the data distribution, and the hash codes are not correlated. Therefore, several learning-based LSH methods were introduced \cite{wang2017survey,thrun2012learning}. They are trained to learn the dataset distribution to improve performance.

Nowadays, due to increasing demand for users' privacy and security, we often need to remove users' data information from Machine Learning (ML) models to satisfy certain privacy and security requirements. During the past ten years, facing an unprecedented scale of ML's application on personal data, many countries have established provisions on the right to be forgotten in data privacy legislation to mandate companies to achieve the erasure of certain personal data when required \cite{bourtoule2021machine}, e.g., the General Data Protection Regulation (GDPR) in the European Union \cite{voigt2017eu}, the California Consumer Privacy Act (CCPA) in the United States \cite{goldman2020introduction}, and the PIPEDA privacy legislation in Canada \cite{privacy}.

This poses the need to have an ANN search algorithm to support fast online data deletion and insertion. However, current learning-based hashing methods require retraining the hash function, which is prohibitive due to the vast time-cost of large-scale data. In this paper, we propose a novel data-distribution-independent hashing method named \textbf{unfolded self-reconstruction locality-sensitive hashing (USR-LSH)} to address the machine unlearning of retrieval problems. To the best of our knowledge, we are the first to address this problem. Empirically, we demonstrate that USR-LSH outperforms the state-of-the-art data-independent LSH in ANN tasks in terms of precision and recall. We also show that USR-LSH has significantly faster data deletion and insertion times than learning-based data-dependent hashing. Finally, we suggest the use of multi-probes and multi-hash tables for future work.

\section{Related Works}

\subsection{SimHash}

Charikar \cite{charikar2002similarity} proposed simHash in the local-sensitive hashing family to generate binary hash code. Given an input vector $\boldsymbol{x} \in \mathbb{R}^d$, the hash function is as follows:
\begin{eqnarray}
    \boldsymbol{y} = sign (\boldsymbol{Wx})
\end{eqnarray}

The $\boldsymbol{W} \in \mathbb{R}^{md}$ is a random projection matrix, where each entry of $\boldsymbol{W}$ is drawn from Gaussian distribution, ${W_{ij}} \sim \mathcal{N}(0,\boldsymbol{
I}_d)$. A row $\boldsymbol{w_i} \in \mathbb{R}^d $ in $\boldsymbol{W}$ can be geometrically viewed as a random unit vector that defines a hyper-plane that is normal to $\boldsymbol{w_i}$ and pass through the origin. We have $m$ such a random hyper-plane in $\boldsymbol{W}$. The $sign(\boldsymbol{Wx})$ slice the input space by $m$ random hyperplanes, and we take the sign depending on which side of the $\boldsymbol{x}$ reside in the hyperplane $\boldsymbol{w_i}$. Hence, this hash function is also called hyper-plane LSH. Since all the hyper-plane pass through the origin, it only preserves cosine distance, but not euclidean distance.

\subsection{Super-bits LSH}

Ji et al.~\cite{ji2012super}  proposed Super-bits LSH (SB-LSH) to improve the random projection hash (simHash). Their follow-up work \cite{ji2014batch} further theoretically proves that, compared to simHash, the Super-bits LSH will provide an unbiased estimate of pairwise angular similarity with a smaller variance for any angle in $(0,\pi)$. Given an input vector $\boldsymbol{x}$ in $\mathbb{R}^d$, the hash function is as follows: 
\begin{eqnarray}
    \boldsymbol{y} = sign (\boldsymbol{Wx}),
\end{eqnarray}
where the W is a stacked random rotation matrix, which is initialized as follows:
\begin{eqnarray}
    \boldsymbol{W} = \sqrt{\frac{d}{h}} [\boldsymbol{R}_1, ..., \boldsymbol{R}_m],
\end{eqnarray}
where $\boldsymbol{R}_i \in \mathbb{R}^{d\times d}$ is a random rotation matrix, where each row in the $\boldsymbol{R}_i$ has unit length  and are orthogonal to each other. We stack the $m$ number of the rotation matrix to produce $\boldsymbol{W}$. To produce each $\boldsymbol{R}_i$, we can produce a Gaussian random matrix $\boldsymbol{S} \in \mathbb{R}^{d\times d} \sim \mathcal{N}(0,\boldsymbol{I}_d)$ where the entries are sampled from the standard Gaussian distribution, then we orthogonalized it to produce a $\boldsymbol{R}_i$. The orthogonalization procedure is the Gram-Schmidt process, which projects the current vector orthogonally onto the orthogonal complement of the subspace spanned by the previous vectors.

\subsection{Faiss Library}

Faiss \cite{johnson2019billion} is a state-of-the-art approximate nearest neighbour (ANN) search library written in highly optimized C++ code with a Python wrapper. In its LSH module \texttt{faiss.IndexLSH}, it hash an input $\boldsymbol{x} \in \mathbb{R}^{d}$ to hash code $\boldsymbol{b} \in \{-1,+1\}^{n}$. When $n<d$, Faiss-LSH uses simHash; when $n>=m$, it uses SB-LSH. Given a query, the \texttt{faiss.IndexLSH} searches for k-nn by doing an exhaustive search over the binary hash code space.

\subsection{Hash Code Ranking}
Given the hash code of dataset $h(\boldsymbol{X})$, and the hash code of query $h(\boldsymbol{q})$, the most basic method to obtain its nearest neighbour is to perform an exhaustive search by computing the hamming distance between $h(\boldsymbol{q})$ and $h(\boldsymbol{X})$, then select the top-k points with the minimum distance as the k-NN.
\\

When data sizes are small, an exhaustive search is often faster than a hash table lookup. This strategy exploits the advantage of binary hash codes that computing hamming distance is a lot more efficient than other distances, such as Euclidean and cosine distance, because it mainly involves bit-wise XOR operation. In this work, we will use hash code ranking to evaluate our LSH algorithm.

\subsection{Hash Table Lookup}

Using hash code ranking, the query speed increase linearly with the size of the dataset. We can achieve constant time complexity with \textbf{Hash Table Lookup} \cite{wang2017survey}. A hash table is an inverted index data structure where the key is hash code $h(\boldsymbol{X})$, and the value is a list of references of data points $\boldsymbol{X}$. Given the hash code of a query $h(\boldsymbol{q})$, \textbf{Hash Table Lookup} is to lookup for the $h(\boldsymbol{q})$ bucket if the bucket exists, the data lying in the bucket $h(\boldsymbol{q})$ are retrieved as the candidates of the nearest neighbours of q.
If the original data is available, this is usually followed by a ranking step to sort the candidates according to their true distances with the query point. Suppose we are querying for k nearest neighbour (k-NN). Then we return the top k candidates with the minimum true distance.
If the original data is not available, DenseFly \cite{sharma2018improving} constructs a low-dimension hash code for hash table lookup and constructs a low-dimension hash code for candidate ranking.

\subsection{Multi-probe}
To guarantee high precision in hash table lookup, the bucket size (i.e., dimension of hash code) needs to be sufficiently large. But this means that the probability of bucket collision will be too low, i.e., $h(\boldsymbol{q})$ often be an empty bucket. To overcome this problem, \cite{lv2007multi} uses the idea that nearest neighbours are more likely to be hashed into the close-by buckets and intelligently probe (lookup) the buckets that are near to $h(\boldsymbol{q})$.
\\

In this work, we focus on constructing the hash function. Besides the hash function, hash table lookup and multi-probe are promising to improve the query speed.  These techniques are orthogonal to our hash function techniques and can be plug-in our method to improve the query speed further. We leave this part as one of our future work.




\section{Methods}
\subsection{Hash function: Unfolded Self-Reconstruction Hashing}

We develop a binary hashing function called Unfolded Self-Reconstruction Hashing locality-sensitive hashing (USR-LSH). Given a data point $\boldsymbol{x} \in \mathbb{R}^d$, our algorithm hashes it into binary space $\boldsymbol{b} \in \{-1,+1\}^{md}$, where the value $md$ is the dimension of the hash code since the hash code is binary, it is also equal to the number of hash bits. In this paper, we use these two terms interchangeably. The $d$ is the dimension of the data point. We can set hyper-parameter $m$ to control the hash code dimension. A higher $m$ will yield better hashing performance (better data reconstruction) but consume more space. \\

\textbf{Self-Reconstruction Objective:} We aim to construct hash codes that better preserve the data information. To achieve this, we minimize the reconstruction error as Eq.(\ref{bObj}):
\begin{align}
\label{bObj}
    \mathcal{J}(\boldsymbol{b}) := ||\boldsymbol{W}\boldsymbol{b} - \boldsymbol{x}||_2^2
\end{align}
where $\boldsymbol{W} \in \mathbb{R}^{d \times md}$ is the random projection matrix constructed by Eq.(\ref{ProjectionW})
\begin{align}
\label{ProjectionW}
    \boldsymbol{W}= \sqrt{\frac{d}{md}} [ \boldsymbol{R}_1, \cdots , \boldsymbol{R}_m],
\end{align}
where $\boldsymbol{R}_i, i \in \{1,\cdots,m\}$ denotes the random orthogonal matrix. The $\boldsymbol{W}$ can be viewed as a decoder matrix that decodes hash code $\boldsymbol{b}$ back to the original data space.
\\

The self-reconstruction objective $\mathcal{J}(\boldsymbol{b})$ is a discrete optimization problem that is challenging to solve.  To alleviate this, we relax the binary constraint ($sign(\cdot)$) by the $tanh(\cdot)$ function. Then,  we  optimize an approximation problem in Eq.(\ref{tObj})
\begin{align}
\label{tObj}
    \mathcal{\widehat{J}}(\boldsymbol{y}) := ||\boldsymbol{W}\boldsymbol{y} - \boldsymbol{x}||_2^2 + \phi_\lambda(\boldsymbol{y}),
\end{align}
with an implicit  regularization $\phi_\lambda(\boldsymbol{y})$ such that $\tanh(\cdot)$ is the proximal function, i.e.,  $\phi_\lambda(\cdot)$ has an efficient proximal operators $g(\cdot) = tanh(\cdot)$ such that $g(z) = \argmin _{x} {\frac{1}{2}(x-z)^2 +  \phi_\lambda (x)} $. \\

\textbf{Unfolded Optimization:} We now show how to optimize the approximation problem $\mathcal{\widehat{J}}(\boldsymbol{y})$. Denotes $y_0 \in \mathbb{R}^{md}$ as the initial hash values and we first initialize its entries with zeros:

\begin{equation}
    \boldsymbol{y_0} = \boldsymbol{0}.
\end{equation}

Given the input $\boldsymbol{x}$ and its hash value $\boldsymbol{y_t} \in \mathbb{R}^{md}$ at time step $t$, the update function $h: (\mathbb{R}^d, \mathbb{R}^{md}) \mapsto \mathbb{R}^{md}$ will produces the hash values of next time step $\boldsymbol{y_{t+1}} \in \mathbb{R}^{md}$:

\begin{equation}
    \begin{aligned}
    \boldsymbol{y_{t+1}} & = h(\boldsymbol{x},\boldsymbol{y_t}),
  \end{aligned}
\end{equation}
where the update function $h$ is given as follows:
\begin{equation}
\label{eqn:hfunc}
    \begin{aligned}
        h(\boldsymbol{x},\boldsymbol{y_t}) = tanh \left(\alpha \boldsymbol{W}^{\top} \boldsymbol{x} + (\boldsymbol{I}-\boldsymbol{W}^{\top}\boldsymbol{W}) \boldsymbol{y_t} \right), 
  \end{aligned}
\end{equation}
where the $\boldsymbol{W}$ is a stacked random rotation matrix and the $\alpha$ is a scalar coefficient.
\\

The $\boldsymbol{W}$ in equation~(\ref{eqn:hfunc}) is initialised as follows:
\begin{eqnarray}
    \boldsymbol{W} = \sqrt{\frac{d}{md}} [\boldsymbol{R}_1, ... , \boldsymbol{R}_m],
\end{eqnarray}
where the $\boldsymbol{R}_i \in \mathbb{R}^{d\times d}$ is a random rotation matrix. The rows of the $\boldsymbol{R}_i$ are orthonormal, i.e., they have unit length one and orthogonal to each other, i.e., $\boldsymbol{R}_i^\top\boldsymbol{R}_i = \boldsymbol{I}_d$. We stack the $m$ number of the rotation matrix to produce $\boldsymbol{W}$. To produce each $\boldsymbol{R}_i$, we can produce a Gaussian random matrix $ \boldsymbol{S} \in \mathbb{R}^{d\times d} \sim \mathcal{N}(0,\boldsymbol{I}_d)$ where the entries are sampled from the standard Gaussian distribution, then we orthogonalized it to produce a $\boldsymbol{R}_i$. In our work, the vector orthogonalization process is done by the single value decomposition (SVD) process.
\\

For the $\alpha$ in Equation~(\ref{eqn:hfunc}), we empirically we find the best value is:
\begin{equation}
    \begin{aligned}
        \alpha = \frac{\sqrt{md}}{2\mathbb{E}_{\boldsymbol{x}\in \boldsymbol{X}}[|\boldsymbol{x}|_2]}, 
  \end{aligned}
\end{equation}
where the $\mathbb{E}_{\boldsymbol{x}\in \boldsymbol{X}}[|\boldsymbol{x}|_2]$ is calculated by taking the average Euclidean norm of the whole dataset $\boldsymbol{X} = \{\boldsymbol{x_0}, \boldsymbol{x_1}, ..., \boldsymbol{x_N}\}$. The purpose is to scale the length of every input point close to one to improve the instance-wise learning. 
\\

We iterate the update function $h$ for $T$ time to produce the final hash value $\boldsymbol{y_T}$. Lastly, the binary hash code $\boldsymbol{b} \in \{-1,+1\}^{md}$ is obtained by the sign of the hash value:
\begin{equation}
    \begin{aligned}
    \boldsymbol{b} = sign(\boldsymbol{y_T}),
  \end{aligned}
\end{equation}
where the sign function is given as:
\begin{equation}
    sign(x) =
    \begin{cases}
      -1 & \text{if } x < 0 \\
      +1  & \text{if }  x \ge 0 
    \end{cases}.
\end{equation}

We unfold the $T$-step update as  USR-LSH  network architecture. The algorithm to optimize $\mathcal{\widehat{J}}(\boldsymbol{y})$ is  given in Algorithm \ref{algo1}. Given a proximal function, \cite{lyu2021neural} theoretically proved that the reconstruction error is reduced, i.e., $\mathcal{\widehat{J}}(\boldsymbol{y}_t) \le \mathcal{\widehat{J}}(\boldsymbol{y}_{t\!-\!1})$, for every iteration $t=\{1, \cdots, T\}$. This work employs the $tanh(\cdot)$ function as the proximal function for an implicit regularization to construct hash codes.  In addition, in practice, the operation can be done in parallel for a batch of the input points $\boldsymbol{X}$.


\begin{algorithm}[t]
  \caption{USR-LSH}
  \label{algo1}
\begin{algorithmic}
    \STATE {\bf Input:} Data $x$, Hash code length $m$, Iteration $T$, scalar coefficient $\alpha$.
    \STATE {\bf Initialize:} $\boldsymbol{y}_0 \leftarrow \textbf{0}$
    \FOR{t= 1 to T }
        \STATE Set $\boldsymbol{y}_{t}  \leftarrow tanh\left(\alpha \boldsymbol{W}^{\top} \boldsymbol{x} + (\boldsymbol{I}-\boldsymbol{W}^{\top}\boldsymbol{W}) \boldsymbol{y}_{t-1} \right)$ 
    \ENDFOR
    \STATE \textbf{Return} $sign(\boldsymbol{y}_T)$
\end{algorithmic}
\end{algorithm}





\subsection{Hash Buckets Probe}

When using hash code ranking, the computational cost increases linearly with the data size. On the other hand, when using hash table lookup, the probability of hash collision becomes too low when the hash code is sufficiently long. To improve on this, \cite{tao2009quality} uses the idea that nearest neighbours are more likely to be hashed into close-by buckets and intelligently probes (looks up) the buckets that are near to $h(\boldsymbol{q})$.
\\

Given a hash code $g(\boldsymbol{q}) = (h_1(\boldsymbol{q}), . . . , h_M(\boldsymbol{q}))$, where $g(\boldsymbol{q}) \in [0, b]^{M}$, Multi-probe LSH \cite{lv2007multi} defines the hash perturbation vector to be $\boldsymbol{\Delta} = (\delta_1, ... , \delta_{M} ) $, where $\delta_i \in \{ -1,0,+1 \} $. Then we apply the perturbation vector to probe the nearby bucket $h(\boldsymbol{q}) + \boldsymbol{\Delta}$. We assign a score to each of the perturbations $\boldsymbol{\Delta_i}$, where smaller scores have a higher probability of yielding points near to $\boldsymbol{q}$. Note that the score of $\boldsymbol{\Delta}$ is a function of both $\boldsymbol{\Delta}$ and the query $\boldsymbol{q}$. We then order the perturbation vectors in increasing order of their (query dependent) scores to generate a sequence of perturbation vectors $( \boldsymbol{\Delta_1}, \boldsymbol{\Delta_2}, ..., \boldsymbol{\Delta_n} )$ to visit $n$ nearby buckets residing at $( h(\boldsymbol{q}) + \boldsymbol{\Delta_1}, h(\boldsymbol{q}) + \boldsymbol{\Delta_2}, ..., h(\boldsymbol{q}) + \boldsymbol{\Delta_n} )$. This is the query-directed probing method described in \cite{lv2007multi}.
\\

In USR-LSH, our hash code is binary, $h(\boldsymbol{q}) \in \{+1,-1\}^{md}$, so instead of adding (or subtracting), our perturbation vector is used to flip the sign of some coordinates of the hash code. We can define our perturbation vector to be $\boldsymbol{\Delta} = (\delta_1, ... , \delta_{M} ) $, where $\delta_i \in \{ -1, +1 \} $. Instead of adding to the hash code, we do element-wise multiplication to flip the sign of the hash code, $h(\boldsymbol{q}) \odot \boldsymbol{\Delta}$. That is, we will probe the nearby bucket in a sequence of $( h(\boldsymbol{q}) \odot \boldsymbol{\Delta_1}, h(\boldsymbol{q}) \odot \boldsymbol{\Delta_2}, ..., h(\boldsymbol{q}) \odot \boldsymbol{\Delta_n} )$.
\\

Our hash value $\boldsymbol{b} \in \mathbb{R}^{md} $ reconstructs the data point in L2 distance, so the closest perturbation is to flip the coordinate that is closest to $0$. Using this idea, we can define the score as the sum of the absolute value of the perturbation coordinates of the code $b$:
\begin{eqnarray}
    score(\boldsymbol{\Delta}) = \sum_{i=0}^{md} ( \frac{1-\delta_i}{2} |b_i|),
\end{eqnarray}
where $\boldsymbol{b_i}$ is the $i$-th coordinate of the hash value $\boldsymbol{b}$. Note that the score of a perturbation vector $\boldsymbol{\Delta}$ depends only on its perturbed coordinates, $\delta_i = -1$. We expect that perturbation vectors with low scores will have $+1$ in most entries.
\\

With this score function, we can generate the probing sequence using a min-heap, same as in \cite{lv2007multi}. We can also easily generalize it to multiple hash tables to further improve precision.

\section{Experiments}

\subsection{Dataset}
We evaluate each algorithm on four datasets as shown in Table~\ref{tab1} from the popular ANN benchmark framework \cite{aumuller2020ann}. We follow \cite{aumuller2020ann} to perform Euclidean normalization on the angular dataset.

\begin{table}
\caption{Dataset used in our experiment.}
\label{tab1}
\centering
\begin{tabular}{|c|c|c|c|}
\hline
Dataset &  Size & Dimension & Distance \\
\hline
Deep1B & 9,990,000 & 96  & Angular   \\
SIFT   & 1,000,000 & 128 & Euclidean \\
GloVe-100 & 1,183,514 & 100 & Angular \\
GloVe-25  & 1,183,514 & 25  & Angular \\
\hline
\end{tabular}
\end{table}

\subsection{Notation and Definitions}

Given a query point $\boldsymbol{q}$, denotes $Y$ as the set of ground truth nearest neighbour (NN) of $\boldsymbol{q}$. Suppose we are querying for $\hat{k}$ nearest neighbour points ($\hat{k}$-query), and denotes the set of query result as $\hat{Y}$, we can assume that the algorithm always returns $\hat{k}$ number of points, i.e., $\hat{k} = |\hat{Y}|$.
\\

The precision@$\hat{k}$ of the $\hat{k}$-query result is defined as the percentage of correct points in the query result. It is defined as follows:

\begin{eqnarray}
    precision@\hat{k} = \frac{|\hat{Y} \cap Y|}{|\hat{Y}|}.
\end{eqnarray}
\\

The recall@$\hat{k}$ of the $\hat{k}$-query result is defined as the percentage of the truth nearest neighbour points $Y$ that are returned in the query result. It is defined as follows:
\begin{eqnarray}
    recall@k = \frac{|\hat{Y} \cap Y|}{|Y|}.
	\label{Eq3}
\end{eqnarray}
\\

In our dataset, we have 10 ground truth points for any query, i.e., $|Y|=10$. The $\hat{k}$ is different in a different experiment. In experiment 1, we query for 10 nearest neighbours, i.e., $\hat{k} = k = 10$. In this case, precision@10 is equal to recall@10. We then evaluate the trade-off between query time and precision@10. In experiment 2, we query for $\hat{k} = \{1,...,100\}$ nearest neighbours, and evaluate the trade-off between precision and recall for each $\hat{k}$. Experiment 3 is to evaluate deletion time which will not be affected by $k$ or $\hat{k}$.

\subsection{Experiment Settings}
Our experiments are based on Martin's ANN-benchmark framework \cite{aumuller2020ann}. All our experiments use full dataset input and evaluate on 1000 query points to produce the average recall and precision. All experiments are running on Intel(R) Xeon(R) Gold 6342 CPU @ 2.80GHz (this CPU has one thread per core). All experiment runs are restricted on one core CPU with one thread by using command \texttt{taskset}. The source code to reproduce our experiment can be found at Anonymous GitHub \footnote{https://anonymous.4open.science/r/ann-benchmarks-3786}.

\subsection{Experiment 1: QPS - Precision@10 trade-off}
In this experiment, we evaluated the trade-off between queries per second (QPS) and precision@10 for each algorithm. Given a query $\boldsymbol{q}$, we obtain its hash code $h(\boldsymbol{q})$, and perform an exhaustive search over dataset hash code $h(X)$ for its $\hat{k}$ nearest neighbours. We do a single query at a time without parallelization.
\\

The ground truth dataset had 10 nearest neighbour points for each query point $\boldsymbol{q}$, with $k=10$, and we queried for 10 nearest neighbours for each query with $\hat{k}=10$. Note that in this case, we set $\hat{k}=k$, so the values of recall and precision are equivalent, i.e., precision@10 = recall@10.
\\

We ran the experiments for each algorithm with different parameter settings and plotted the Pareto front of each algorithm. We set the parameters such that every algorithm had an equivalent number of hash bits. For example, if we set $m$ for simHash and USR-LSH, then we would set the SB-LSH parameter to $\texttt{n\_bits}=md$, where $d$ is the data dimension so that the number of hash bits is equivalent. Table \ref{tab2} shows the parameters used in this experiment.
\\

\begin{figure}[htbp]
\begin{subfigure}{.48\linewidth}
\centering
\includegraphics[width=\linewidth]{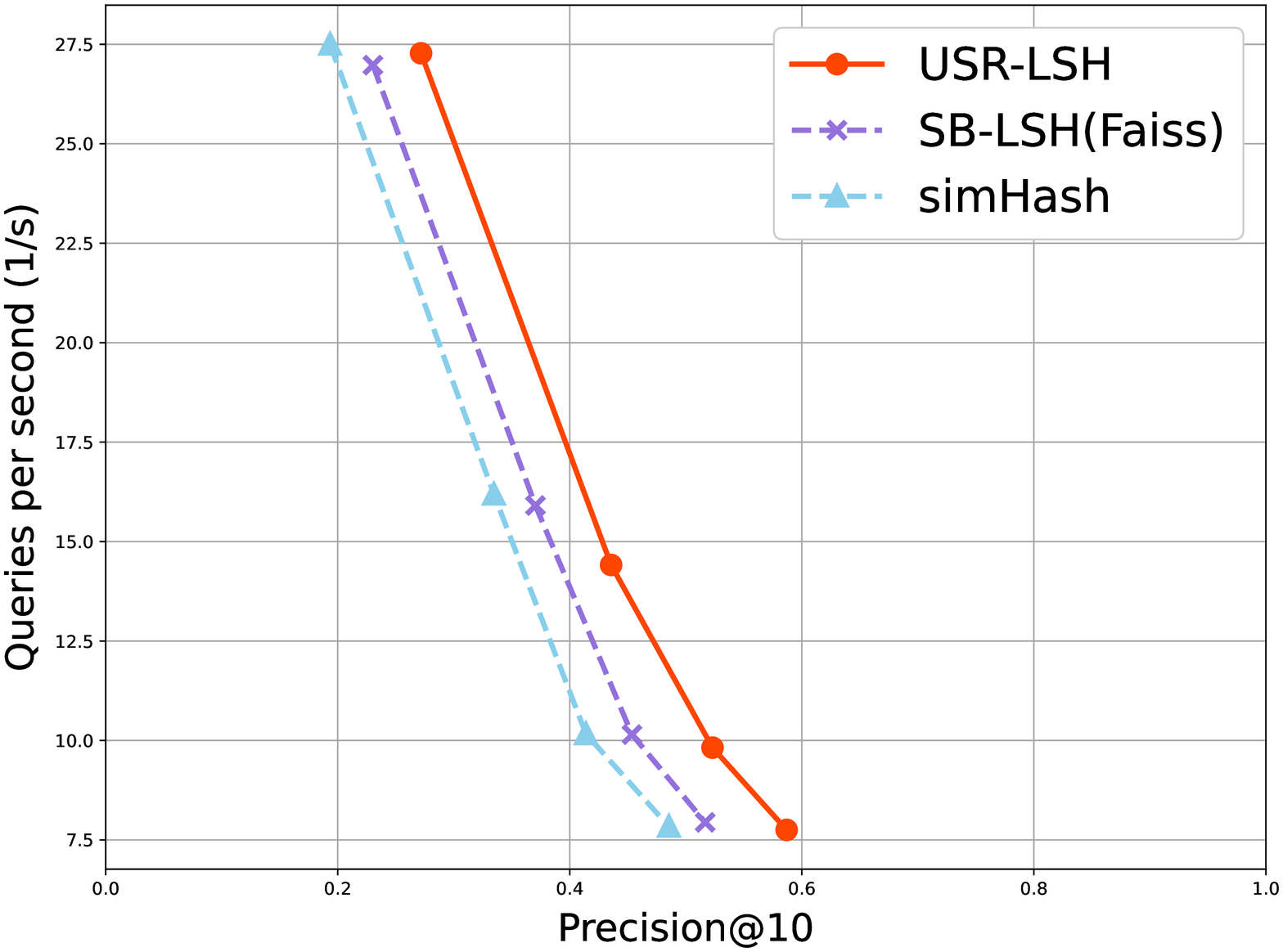}
\caption{Deep1B}\label{exp1a}
\end{subfigure}
\hfill
\begin{subfigure}{.48\linewidth}
\centering
\includegraphics[width=\linewidth]{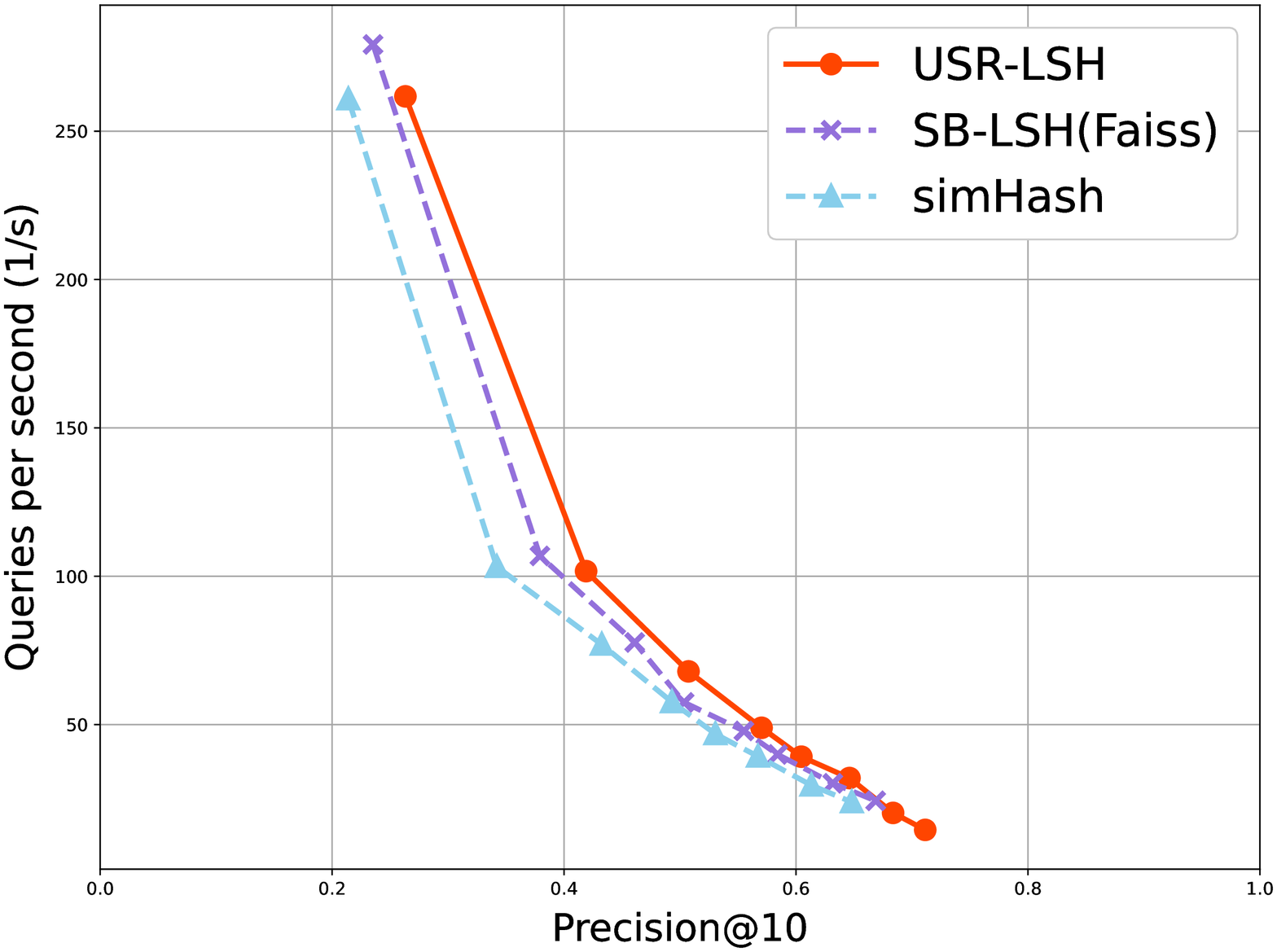}
\caption{SIFT}\label{exp1b}
\end{subfigure}
\begin{subfigure}{.48\linewidth}
\centering
\includegraphics[width=\linewidth]{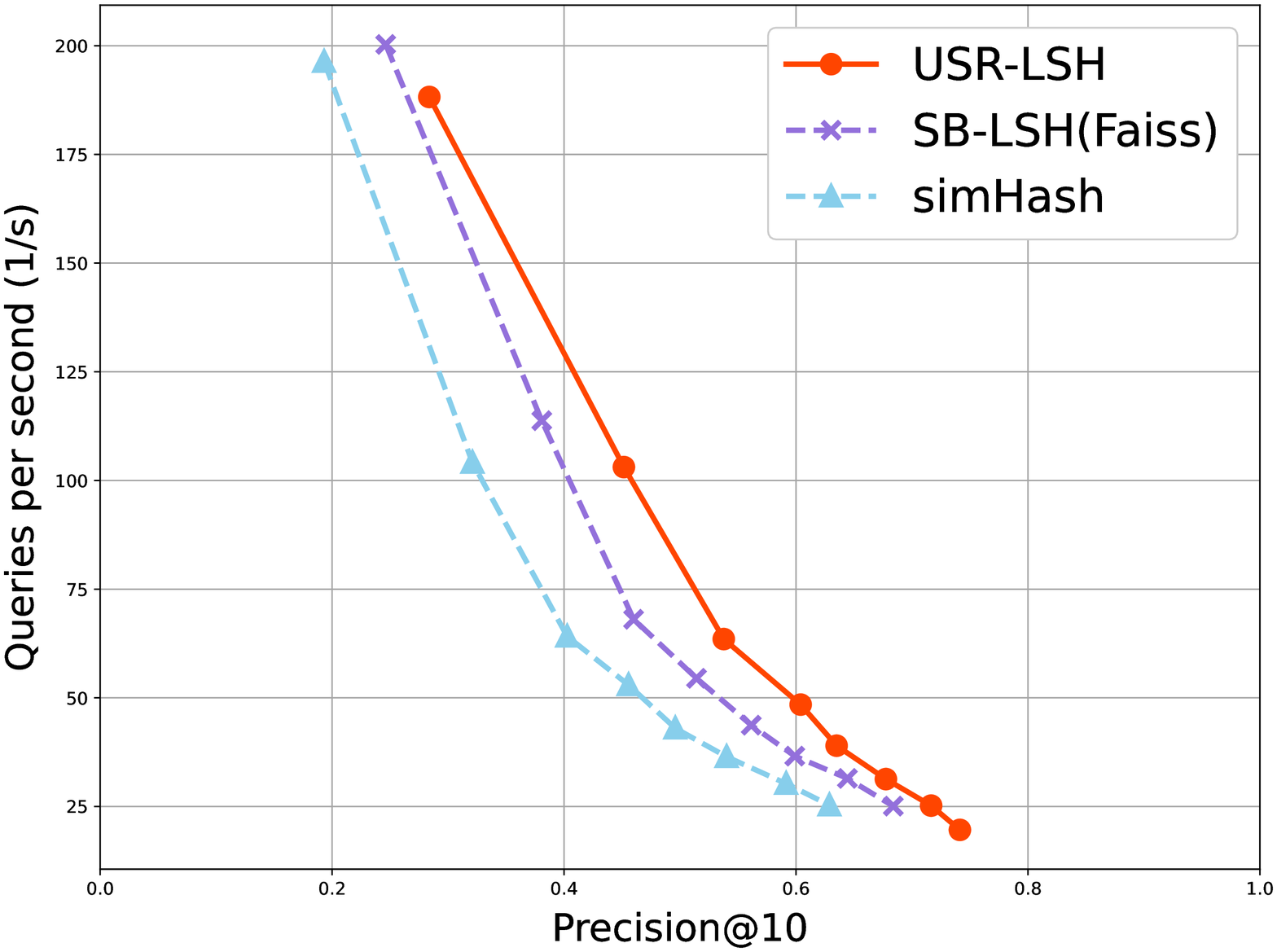}
\caption{Glove-100}\label{exp1c}
\end{subfigure}
\hfill
\begin{subfigure}{.48\linewidth}
\centering
\includegraphics[width=\linewidth]{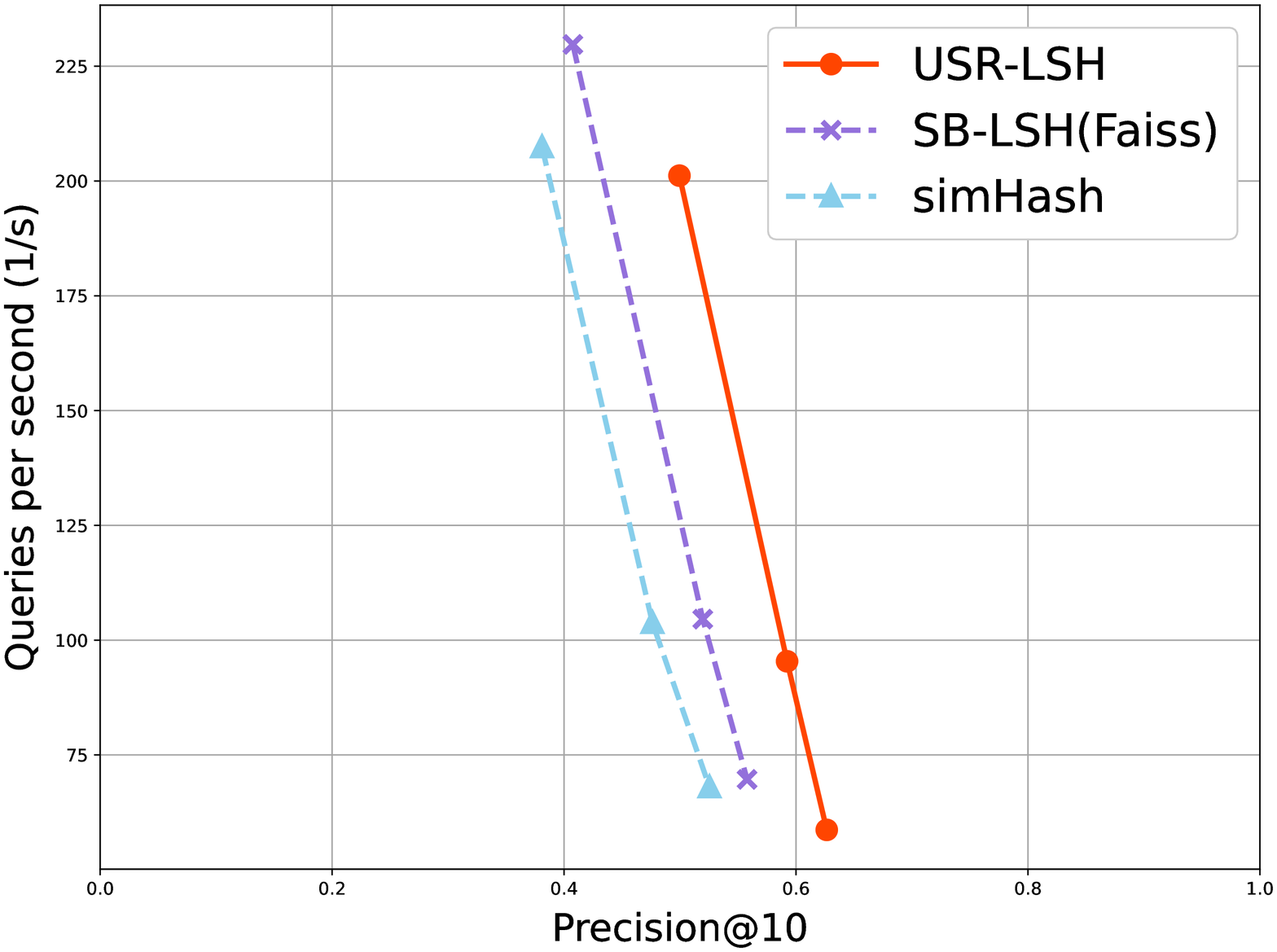}
\caption{Glove-25}\label{exp1d}
\end{subfigure}
\caption{Trade off between Queries per second (1/s) and Precision@10}
\label{fig1}
\end{figure}

Fig. \ref{fig1} shows the QPS and precision@10 plot of each dataset. Our USR-LSH outperformed simHash and SB-LSH on all datasets. We can see that USR-LSH has a slightly slower query time than the others because it iterates for $T$ times, but the improvement of precision makes the Pareto front beyond the simHash and SB-LSH.
\\

Note that SB-LSH is implemented in the official Faiss library, which is written in highly optimized C++ code. For simHash and (our) USR-LSH, the hash function is implemented in Python and pre-compiled with JAX's just-in-time compiler \cite{jax2018github}. Then the hash value is passed to the Faiss library for packing bits and hash code hamming ranking for the k nearest neighbours. If the USR-LSH is further optimized with C++ code, we can expect faster query speed.

\begin{table}[htbp]

\caption{All algorithm parameters are used in Experiment 1 for each Dataset. The lowest and highest (underlined) parameters are used in Experiment 2.}
\centering
\label{tab2}
\resizebox{\textwidth}{!}{
\begin{tabular}{|c|ccc|}

\hline
\multirow{2}{*}{Dataset} & \multicolumn{3}{c|}{Parameters}                                                                                                        \\ \cline{2-4} 
                         & \multicolumn{1}{c|}{simHash (m=)}            & \multicolumn{1}{c|}{SB-LSH (n\_bits=)}                        & USR-LSH (T=17,m=)       \\ \hline
Deep1B                   & \multicolumn{1}{c|}{\{\underline{2},3,4,5,6,7,\underline{8}\}}       & \multicolumn{1}{c|}{\{\underline{192},288,384,480,576,672,\underline{768}\}}          & \{\underline{2},3,4,5,6,7,\underline{8}\}       \\ \hline
SIFT                     & \multicolumn{1}{c|}{\{\underline{2},4,6,8,10,12,16,\underline{20}\}} & \multicolumn{1}{c|}{\{\underline{256},512,768,1024,1280,1536,2048,\underline{2560}\}} & \{\underline{2},4,6,8,10,12,16,\underline{20}\} \\ \hline
GloVe-100                & \multicolumn{1}{c|}{\{\underline{2},4,6,8,10,12,16,\underline{20}\}} & \multicolumn{1}{c|}{\{\underline{200},400,600,800,1000,1200,1600,\underline{2000}\}}  & \{\underline{2},4,6,8,10,12,16,\underline{20}\} \\ \hline
GloVe-25                 & \multicolumn{1}{c|}{\{\underline{2},4,6,8,10,12,16,\underline{20}\}} & \multicolumn{1}{c|}{\{\underline{50},100,150,200,250,300,400,\underline{500}\}}       & \{\underline{2},4,6,8,10,12,16,\underline{20}\} \\ \hline

\end{tabular}}
\end{table}

\subsection{Experiment 2: Precision-Recall trade-off}
In this experiment, we evaluate the precision-recall trade-off for each algorithm. The number of ground truth nearest neighbours is $|Y|=k=10$. Given a query point $\boldsymbol{q}$, we search for its 1 to 100 nearest neighbours, ie, perform $\hat{k}$-query for $\hat{k}={1, ..., 100}$. Then, we evaluate the precision and recall metrics of each $\hat{k}$-query to produce the precision-recall (PR) curve. Lastly, we calculate the area under the PR curve (PR-AUC) to evaluate the performance, the higher PR-AUC means the better performance.
\\

We ran the experiments on all datasets (Table \ref{tab1}). We evaluate the algorithm performance in low and high hash code dimension settings (the parameters underlined in Table \ref{tab2}). Same as Experiment 1, these parameters are set such that the algorithms are compared with the same number of hash code.
\\

\begin{figure}[htbp]
\centering
\begin{subfigure}{.45\linewidth}
\includegraphics[width=\linewidth]{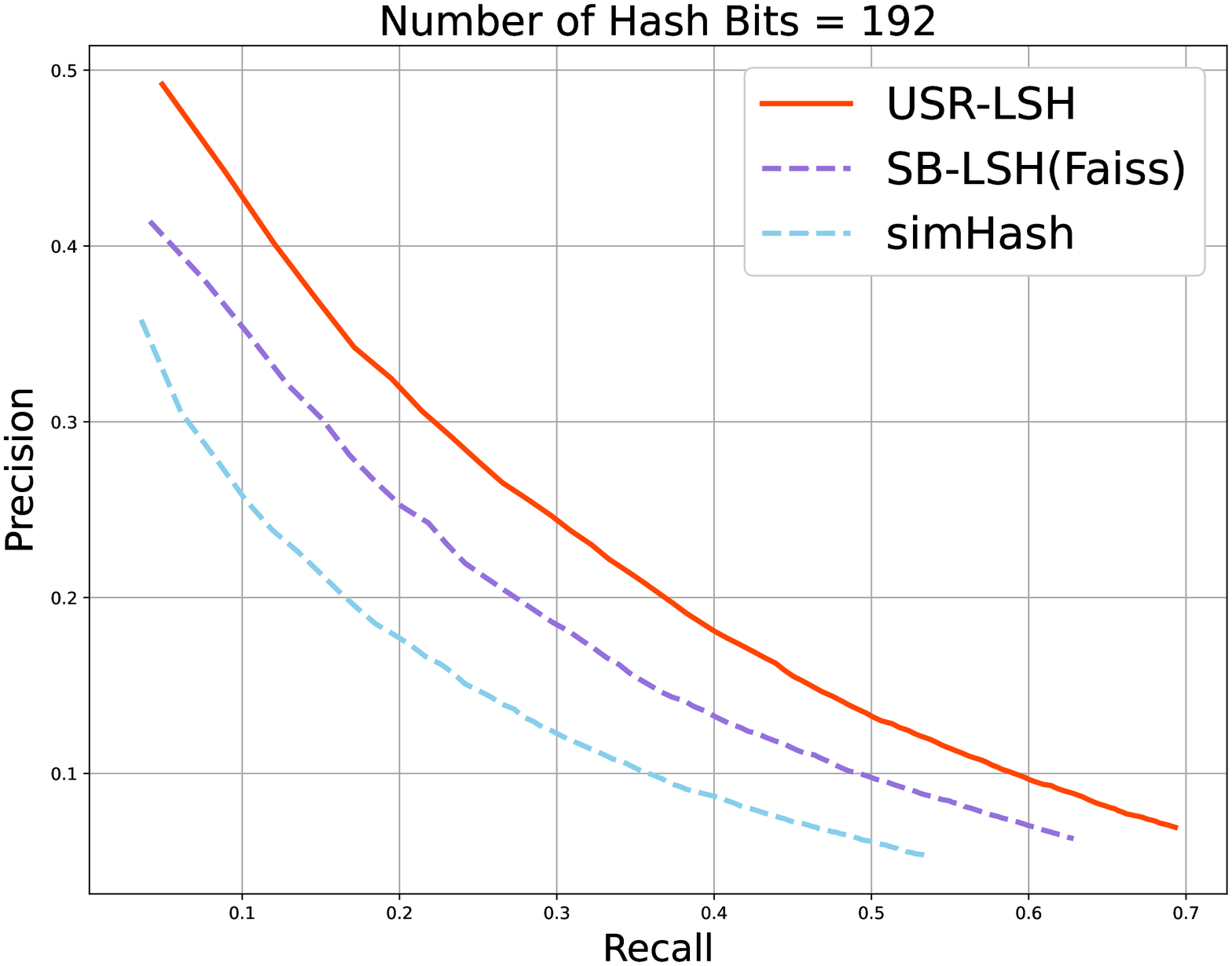}
\caption{Deep1B}
\end{subfigure}
\hfill
\begin{subfigure}{.45\linewidth}
\includegraphics[width=\linewidth]{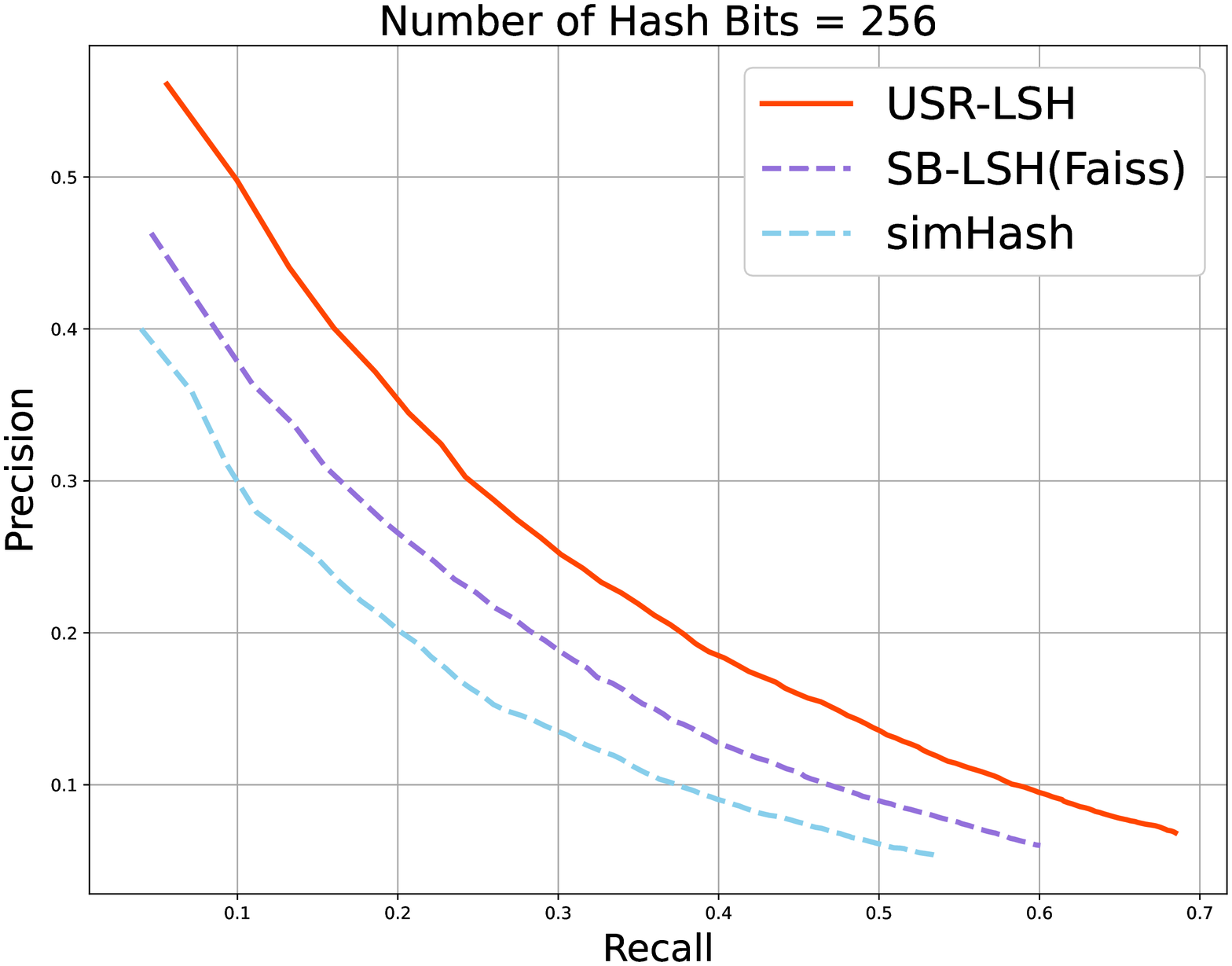}
\caption{SIFT}
\end{subfigure}
\begin{subfigure}{.45\linewidth}
\includegraphics[width=\linewidth]{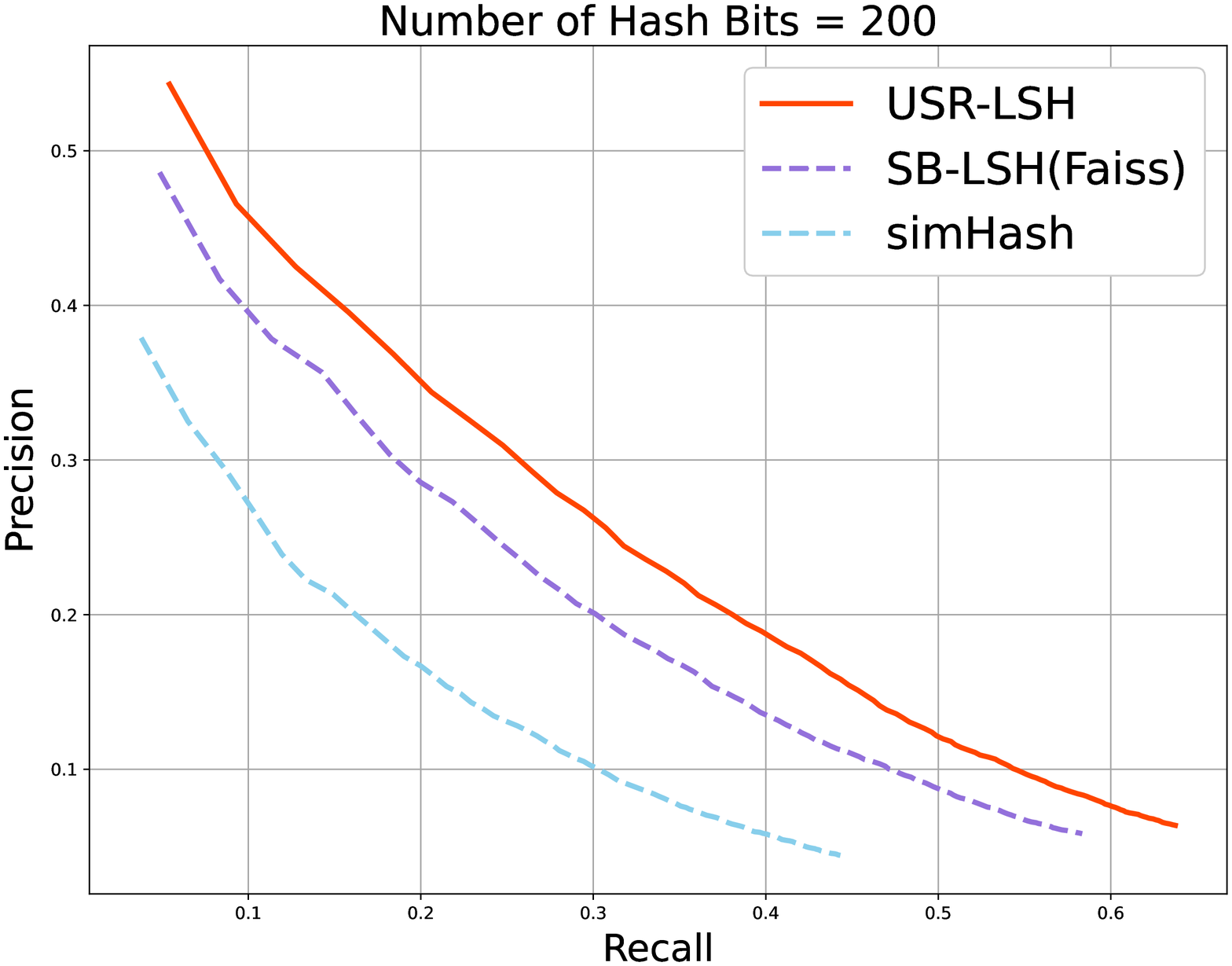}
\caption{Glove-100}
\end{subfigure}
\hfill
\begin{subfigure}{.45\linewidth}
\includegraphics[width=\linewidth]{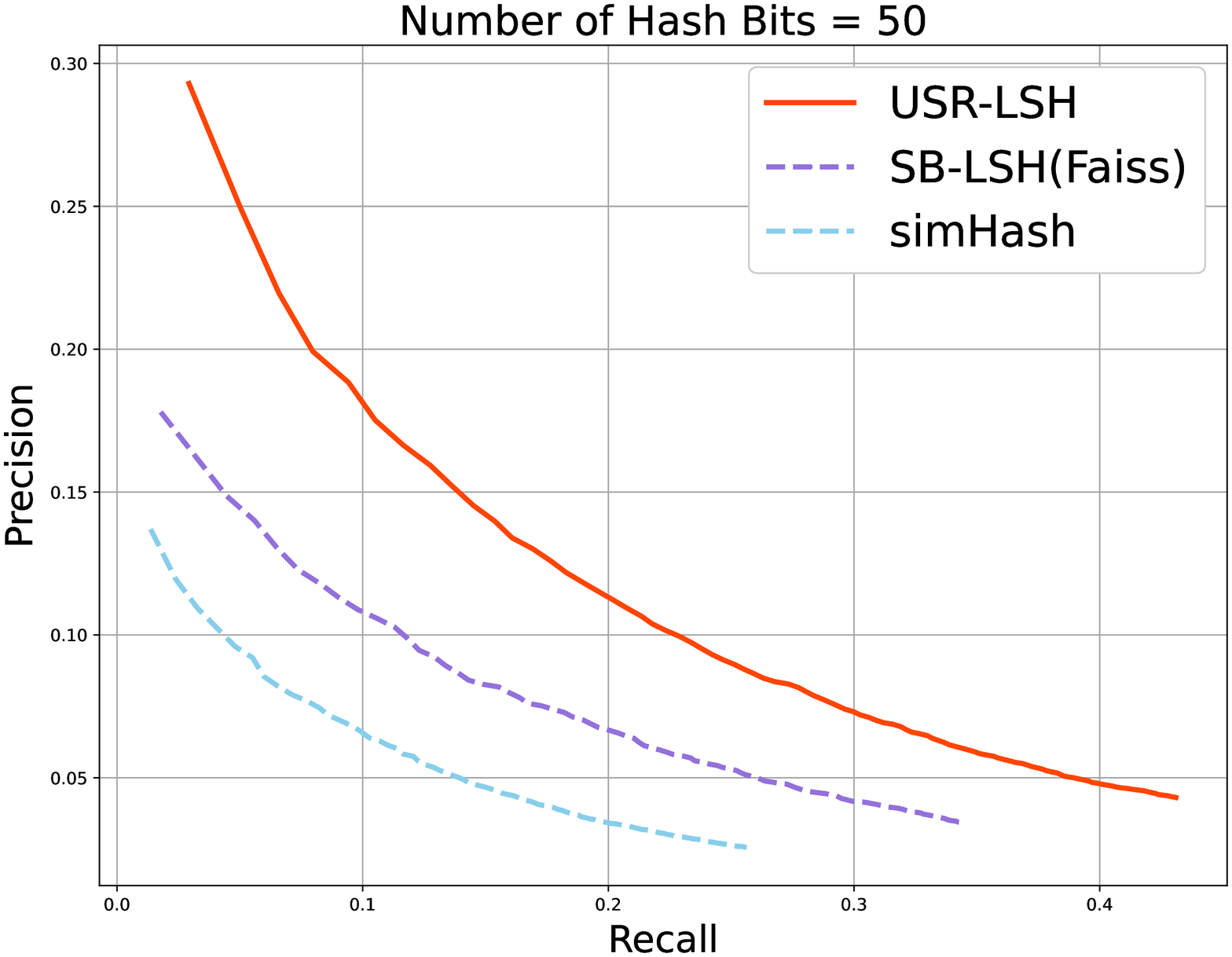}
\caption{Glove-25}
\end{subfigure}
\caption{Precision and Recall curve with low dimensional hash code setting.}
\label{fig:prLow}

\begin{subfigure}{.45\linewidth}
\includegraphics[width=\linewidth]{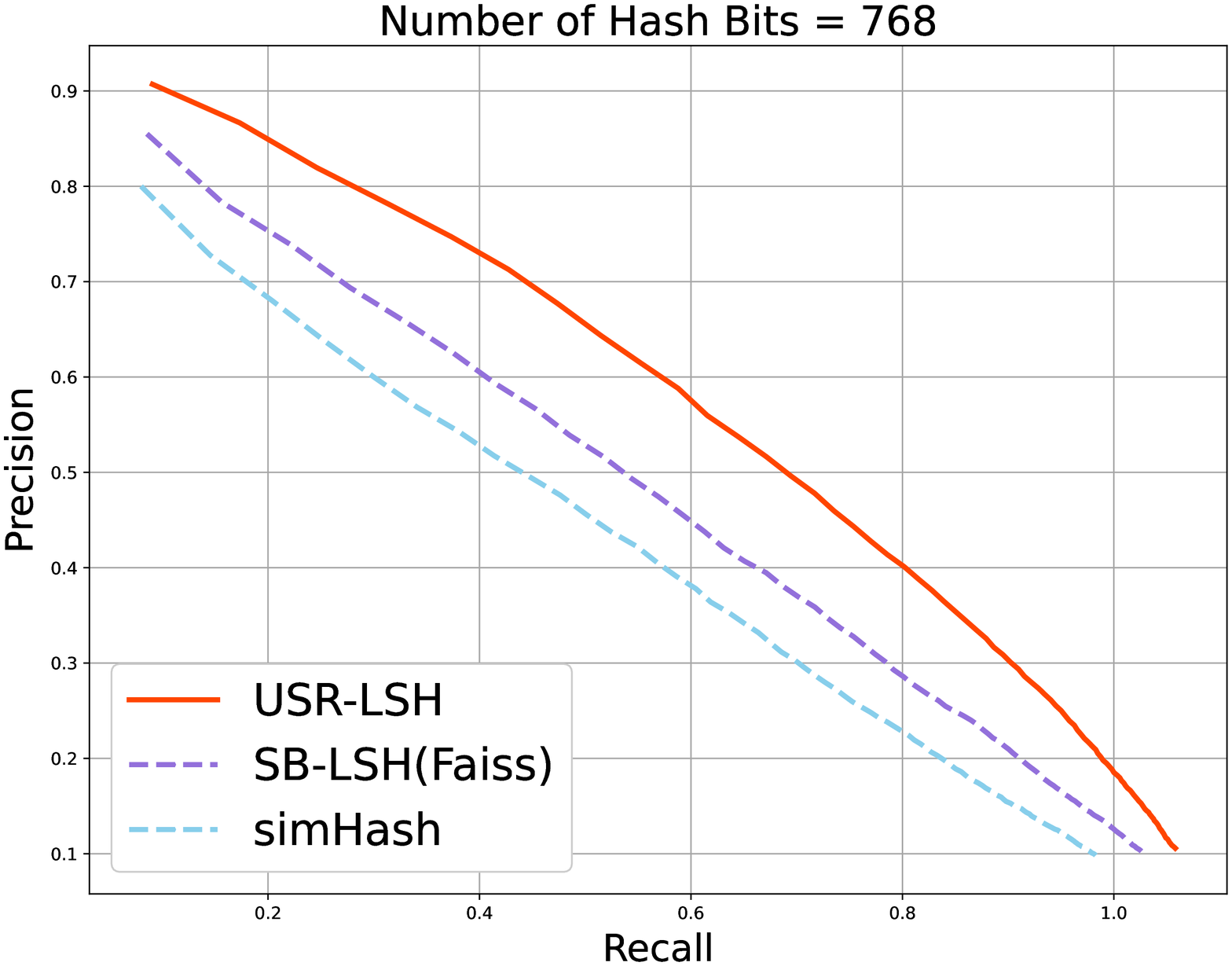}
\caption{Deep1B}
\end{subfigure}
\hfill
\begin{subfigure}{.45\linewidth}
\includegraphics[width=\linewidth]{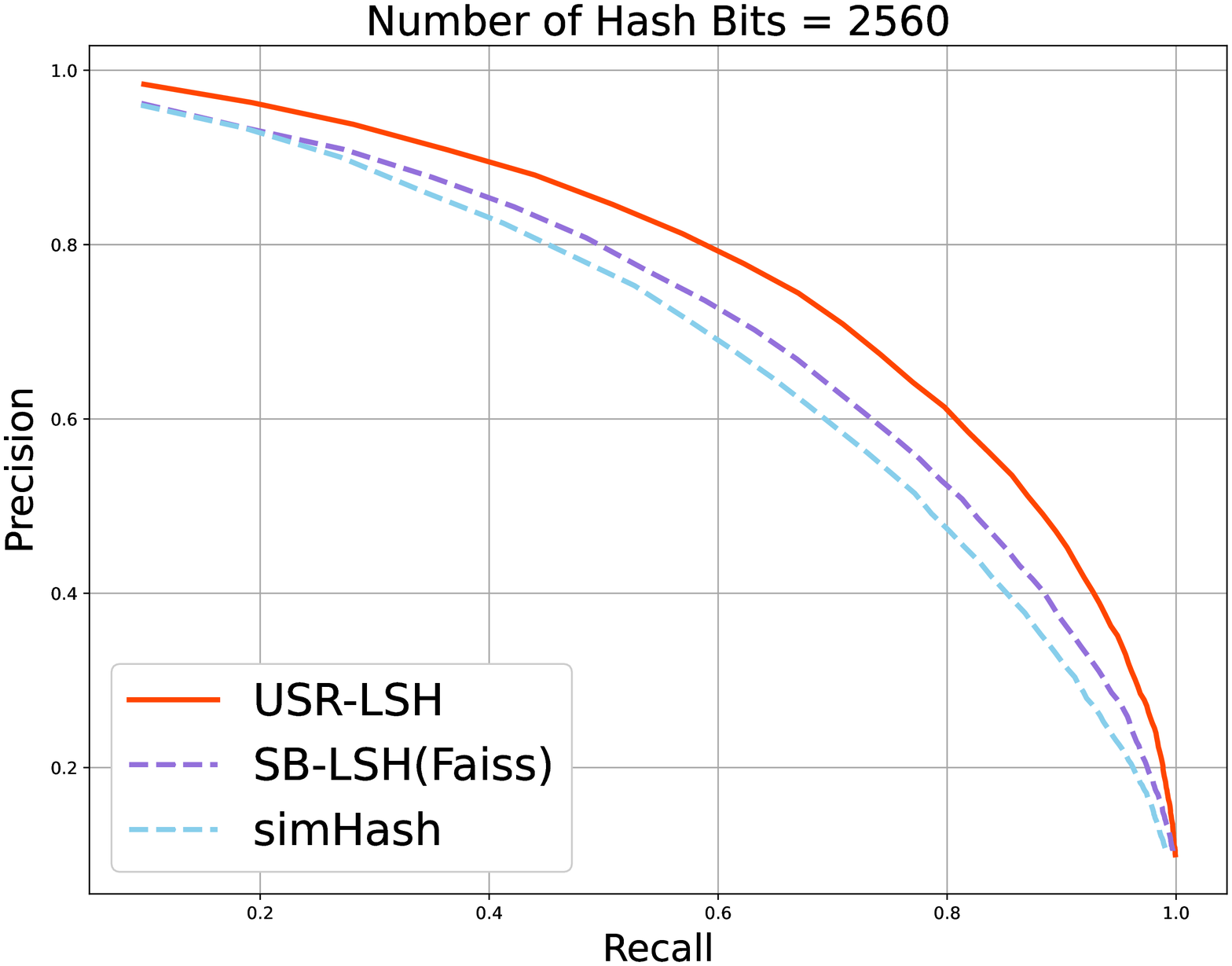}
\caption{SIFT}
\end{subfigure}
\begin{subfigure}{.45\linewidth}
\includegraphics[width=\linewidth]{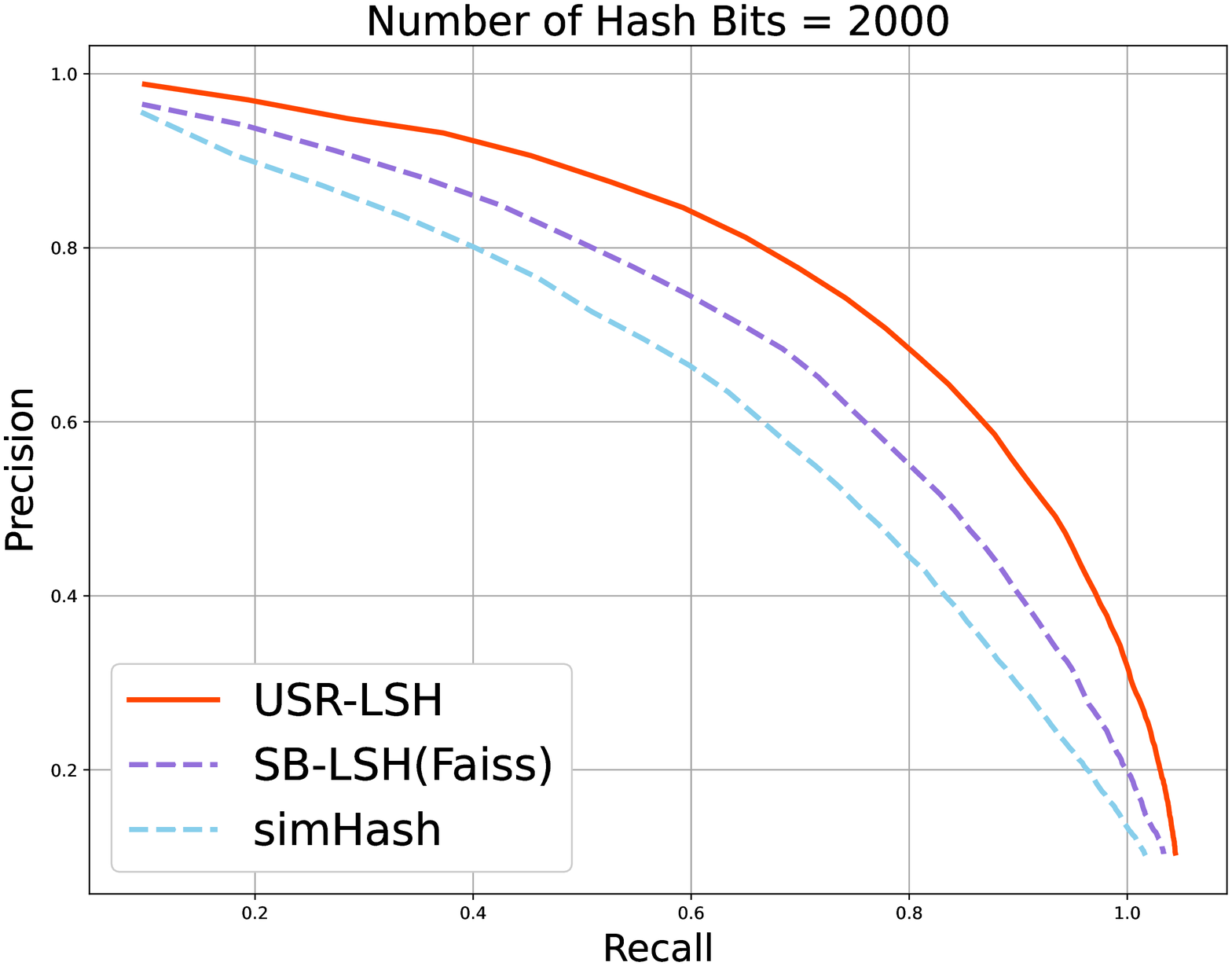}
\caption{Glove-100}
\end{subfigure}
\hfill
\begin{subfigure}{.45\linewidth}
\includegraphics[width=\linewidth]{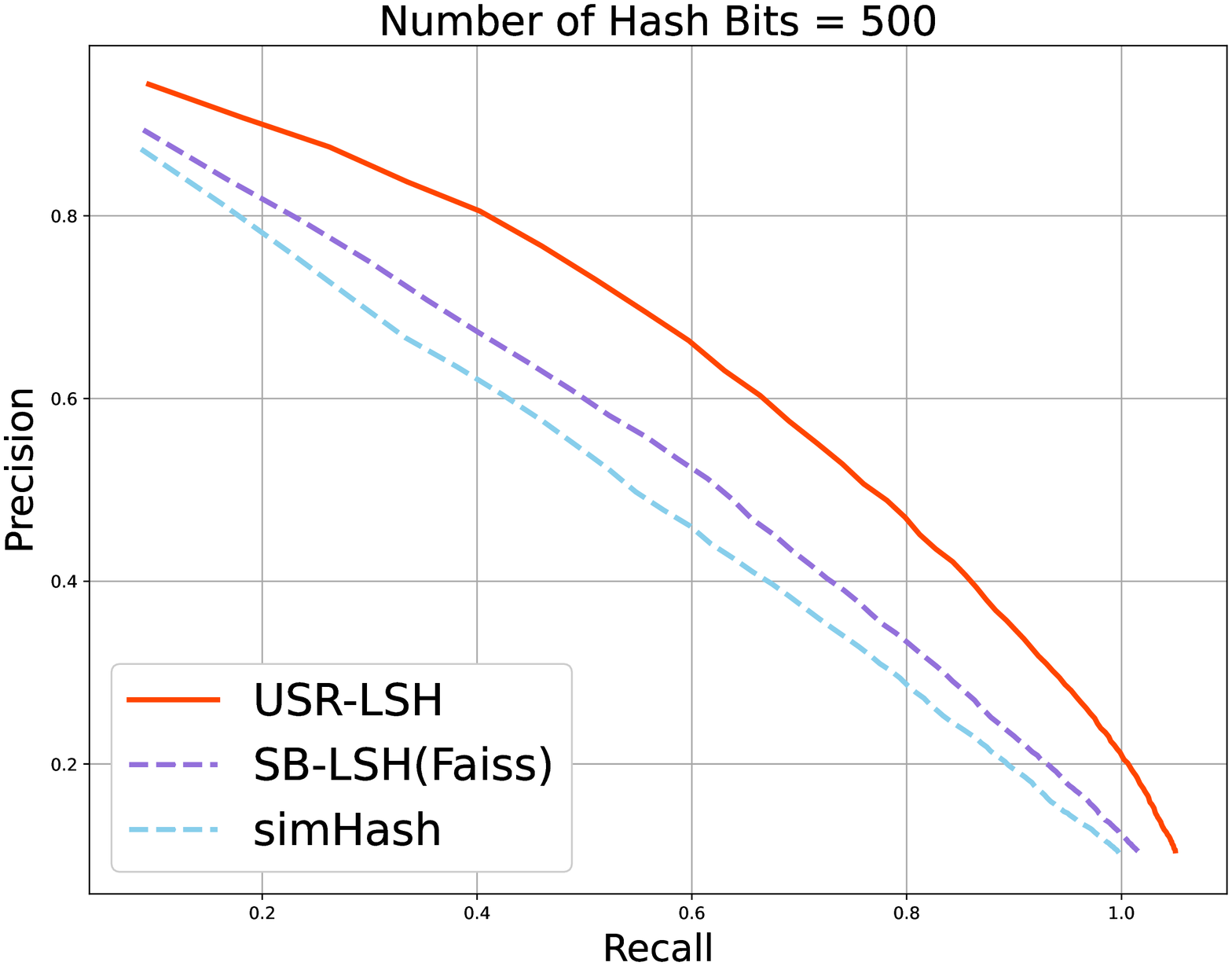}
\caption{Glove-25}
\end{subfigure}
\caption{Precision and Recall curve with high dimensional hash code setting.}
\label{fig:prHigh}
\end{figure}

Fig. \ref{fig:prLow} and Fig. \ref{fig:prHigh} show the PR-curve of each algorithm in all datasets with low and high dimensional hash codes, respectively. We can observe that our USR-LSH outperform the others algorithm in every dataset with both low and high hash code setting.
\\

Table \ref{tab3} shows the area under the PR curves. We use simHash as a baseline and compare the AUC improvement (in percentage) of SB-LSH and our USR-LSH, for each dataset. On average, our USR-LSH improves over 131\% in low dimension hash code setting, and 28\% in high dimension hash code setting. When compared to SB-LSH, USR-LSH achieves more than double improvement (+62\% $\rightarrow$ +131\%) with low-dimension hash code and almost triple improvement (+10\% $\rightarrow$ +28\%) with high-dimension hash code.

\begin{table}[htbp]
\caption{PR-AUC comparison with simHash baseline.}
\label{tab3}
\centering
\begin{tabular}{|c|c|cccc|c|}
\hline
\multirow{2}{*}{Algorithm} & \multirow{2}{*}{Hash Dim} & \multicolumn{4}{c|}{Datasets}                                                                                          & \multirow{2}{*}{Avg Improve} \\ \cline{3-6}
                           &                           & \multicolumn{1}{c|}{Deep1B}      & \multicolumn{1}{c|}{SIFT}        & \multicolumn{1}{c|}{Glove-100}    & Glove-25     &                              \\ \hline
\multirow{2}{*}{USR-LSH}   & low                       & \multicolumn{1}{c|}{0.14(+92\%)} & \multicolumn{1}{c|}{0.15(+82\%)} & \multicolumn{1}{c|}{0.14(+123\%)} & 0.05(+225\%) & \textbf{+131\%}              \\ \cline{2-7} 
                           & high                      & \multicolumn{1}{c|}{0.55(+41\%)} & \multicolumn{1}{c|}{0.68(+13\%)} & \multicolumn{1}{c|}{0.74(+25\%)}  & 0.61(+33\%)  & \textbf{+28\%}               \\ \hline
\multirow{2}{*}{SB-LSH}    & low                       & \multicolumn{1}{c|}{0.11(+49\%)} & \multicolumn{1}{c|}{0.11(+34\%)} & \multicolumn{1}{c|}{0.11(+78\%)}  & 0.03(+86\%)  & +62\%                        \\ \cline{2-7} 
                           & high                      & \multicolumn{1}{c|}{0.45(+16\%)} & \multicolumn{1}{c|}{0.63(+4\%)}  & \multicolumn{1}{c|}{0.66(+12\%)}  & 0.50(+10\%)  & +10\%                        \\ \hline
\multirow{2}{*}{simHash}   & low                       & \multicolumn{1}{c|}{0.08}        & \multicolumn{1}{c|}{0.08}        & \multicolumn{1}{c|}{0.06}         & 0.01         & (baseline)                   \\ \cline{2-7} 
                           & high                      & \multicolumn{1}{c|}{0.39}        & \multicolumn{1}{c|}{0.61}        & \multicolumn{1}{c|}{0.59}         & 0.45         & (baseline)            \\ \hline       
\end{tabular}
\end{table}

\begin{figure}[htbp]
\centering
\includegraphics[width=0.8\textwidth]{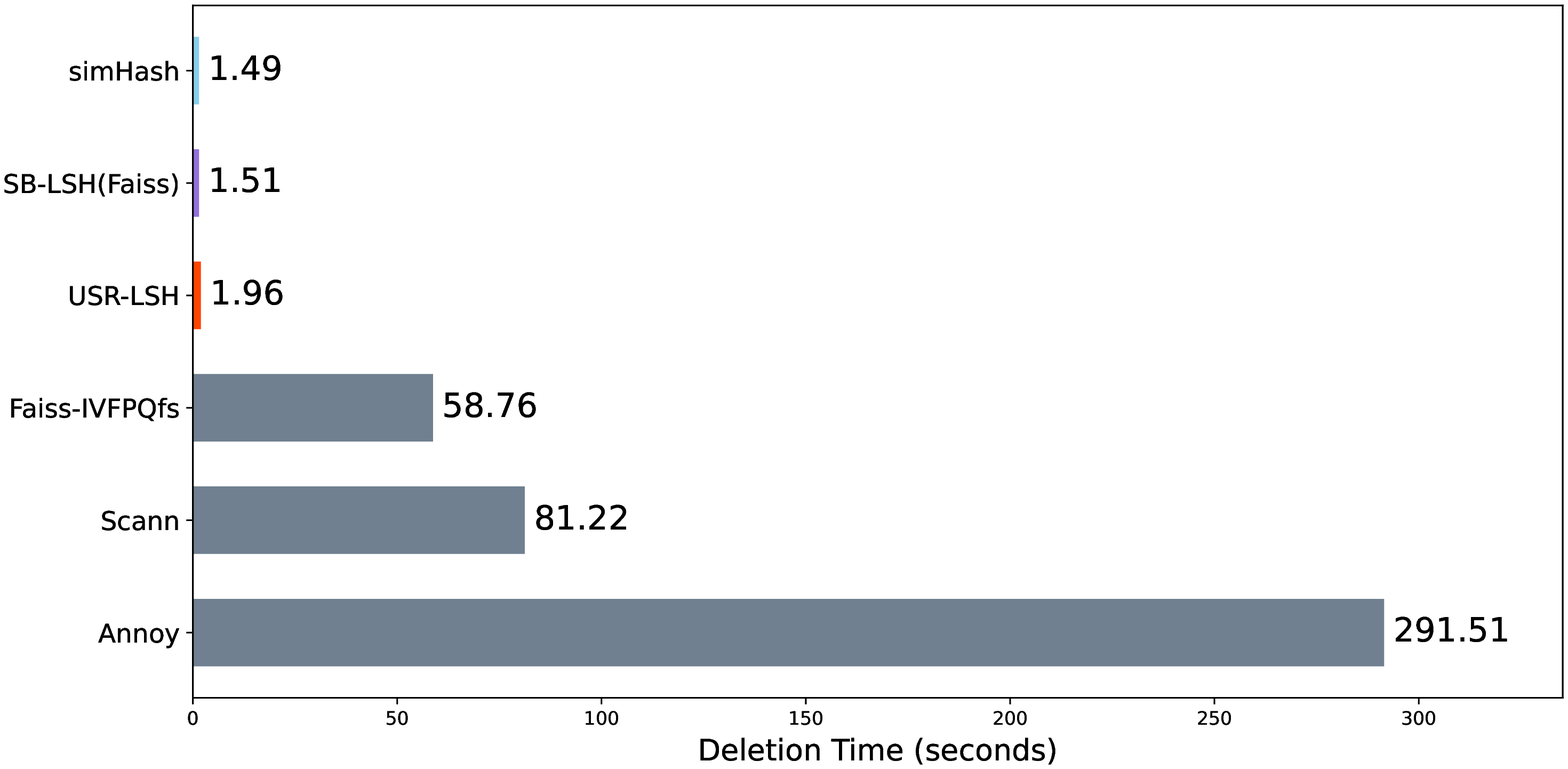}
\caption{Average Deletion time (seconds).} \label{fig3}
\end{figure}

\subsection{Experiment 3: Data Deletion Time}
In this experiment, we compare the data deletion time with some popular data-dependent ANN search algorithms: Scann \cite{avq_2020}, Annoy \cite{annoy}, and the best-performing module in Faiss, Faiss-IVFPQfs \cite{johnson2019billion}. We delete 10 data points, one point at a time, and measure the average deletion time. For Scann, Annoy, and Faiss-IVFPQfs, we retrain the dataset after each deletion. For our USR-LSH, we recalculate the average Euclidean norm $\mathbb{E}_{\boldsymbol{x}\in \boldsymbol{X}}[|\boldsymbol{x}|_2]$ of the dataset for setting coefficient $\alpha$, after each deletion.
\\

We run the experiment on our biggest dataset, Deep1B. The data-dependent approach usually has better performance, so we pick the lowest parameter settings from \cite{aumuller2020ann} (which yield the fastest training time) for a fair comparison. For simHash, SB-LSH, and USR-LSH, we set the highest parameter setting of Deep1B from Table \ref{tab2} (ie, hash bits = 768).
\\

Fig. \ref{fig3} shows the average deletion time (in seconds) of each algorithm. Observed that data-independent algorithms (simHash, SB-LSH, and USR-LSH) can delete significantly faster than the data-dependent algorithms (Scann, Annoy, and Faiss-IVFPQfs). Our USR-LSH is slightly slower than simHash and SB-LSH due to the calculation of the dataset's average L2 norm, but this is negligible when compared to the retraining time of the data-dependent algorithms.

\section{Conclusion}
In this paper, we introduce a novel data-distribution-dependent hashing method named \textbf{unfolded self-reconstruction locality-sensitive hashing (USR-LSH)} to address the problem of machine unlearning in retrieval tasks. This algorithm is supported by the theoretical work presented in \cite{lyu2021neural} We empirically demonstrate that it outperforms state-of-the-art data-independent LSH in terms of precision and recall, and has superior data deletion time compared to other data-dependent ANN algorithms. We also suggest using multi-probes and multi-hash tables for future work.

\section{Ethical Statement}
The dataset used in this work is publicly available without personal information. Our work present a technique for fast neighbour search. It does not have direct impact for policing and military area.

\bibliographystyle{splncs04}
\bibliography{USRLSH}

\begin{thebibliography}{10}
\providecommand{\url}[1]{\texttt{#1}}
\providecommand{\urlprefix}{URL }
\providecommand{\doi}[1]{https://doi.org/#1}

\bibitem{aumuller2020ann}
Aum{\"u}ller, M., Bernhardsson, E., Faithfull, A.: Ann-benchmarks: A
  benchmarking tool for approximate nearest neighbor algorithms. Information
  Systems  \textbf{87},  101374 (2020)

\bibitem{bentley1975multidimensional}
Bentley, J.L.: Multidimensional binary search trees used for associative
  searching. Communications of the ACM  \textbf{18}(9),  509--517 (1975)

\bibitem{annoy}
Bernhardsson, E.: Annoy: Approximate nearest neighbors oh yeah (2017),
  \url{https://github.com/spotify/annoy}

\bibitem{bourtoule2021machine}
Bourtoule, L., Chandrasekaran, V., Choquette-Choo, C.A., Jia, H., Travers, A.,
  Zhang, B., Lie, D., Papernot, N.: Machine unlearning. In: 2021 IEEE Symposium
  on Security and Privacy (SP). pp. 141--159. IEEE (2021)

\bibitem{jax2018github}
Bradbury, J., Frostig, R., Hawkins, P., Johnson, M.J., Leary, C., Maclaurin,
  D., Necula, G., Paszke, A., Vander{P}las, J., Wanderman-{M}ilne, S., Zhang,
  Q.: {JAX}: composable transformations of {P}ython+{N}um{P}y programs (2018),
  \url{http://github.com/google/jax}

\bibitem{privacy}
of~the Privacy Commissioner~of Canada, O.: Announcement: Privacy commissioner
  seeks federal court determination on key issue for canadians\' online
  reputation,
  \url{https://www.priv.gc.ca/en/opc-news/news-and-announcements/2018/an_181010}

\bibitem{charikar2002similarity}
Charikar, M.S.: Similarity estimation techniques from rounding algorithms. In:
  Proceedings of the thiry-fourth annual ACM symposium on Theory of computing.
  pp. 380--388 (2002)

\bibitem{gionis1999similarity}
Gionis, A., Indyk, P., Motwani, R., et~al.: Similarity search in high
  dimensions via hashing. In: Vldb. vol.~99, pp. 518--529 (1999)

\bibitem{goldman2020introduction}
Goldman, E.: An introduction to the california consumer privacy act (ccpa).
  Santa Clara Univ. Legal Studies Research Paper  (2020)

\bibitem{avq_2020}
Guo, R., Sun, P., Lindgren, E., Geng, Q., Simcha, D., Chern, F., Kumar, S.:
  Accelerating large-scale inference with anisotropic vector quantization. In:
  International Conference on Machine Learning (2020),
  \url{https://arxiv.org/abs/1908.10396}

\bibitem{indyk1998approximate}
Indyk, P., Motwani, R.: Approximate nearest neighbors: towards removing the
  curse of dimensionality. In: Proceedings of the thirtieth annual ACM
  symposium on Theory of computing. pp. 604--613 (1998)

\bibitem{ji2012super}
Ji, J., Li, J., Yan, S., Zhang, B., Tian, Q.: Super-bit locality-sensitive
  hashing. Advances in neural information processing systems  \textbf{25}
  (2012)

\bibitem{ji2014batch}
Ji, J., Yan, S., Li, J., Gao, G., Tian, Q., Zhang, B.: Batch-orthogonal
  locality-sensitive hashing for angular similarity. IEEE transactions on
  pattern analysis and machine intelligence  \textbf{36}(10),  1963--1974
  (2014)

\bibitem{johnson2019billion}
Johnson, J., Douze, M., J{\'e}gou, H.: Billion-scale similarity search with
  {GPUs}. IEEE Transactions on Big Data  \textbf{7}(3),  535--547 (2019)

\bibitem{lv2007multi}
Lv, Q., Josephson, W., Wang, Z., Charikar, M., Li, K.: Multi-probe lsh:
  efficient indexing for high-dimensional similarity search. In: Proceedings of
  the 33rd international conference on Very large data bases. pp. 950--961
  (2007)

\bibitem{lyu2021neural}
Lyu, Y., Tsang, I.: Neural optimization kernel: Towards robust deep learning.
  arXiv preprint arXiv:2106.06097  (2021)

\bibitem{sharma2018improving}
Sharma, J., Navlakha, S.: Improving similarity search with high-dimensional
  locality-sensitive hashing. arXiv preprint arXiv:1812.01844  (2018)

\bibitem{tao2009quality}
Tao, Y., Yi, K., Sheng, C., Kalnis, P.: Quality and efficiency in high
  dimensional nearest neighbor search. In: Proceedings of the 2009 ACM SIGMOD
  International Conference on Management of data. pp. 563--576 (2009)

\bibitem{thrun2012learning}
Thrun, S., Pratt, L.: Learning to learn. Springer Science \& Business Media
  (2012)

\bibitem{voigt2017eu}
Voigt, P., Von~dem Bussche, A.: The eu general data protection regulation
  (gdpr). A Practical Guide, 1st Ed., Cham: Springer International Publishing
  \textbf{10}(3152676),  10--5555 (2017)

\bibitem{wang2014hashing}
Wang, J., Shen, H.T., Song, J., Ji, J.: Hashing for similarity search: A
  survey. arXiv preprint arXiv:1408.2927  (2014)

\bibitem{wang2017survey}
Wang, J., Zhang, T., Sebe, N., Shen, H.T., et~al.: A survey on learning to
  hash. IEEE transactions on pattern analysis and machine intelligence
  \textbf{40}(4),  769--790 (2017)

\end{thebibliography}

\end{document}